\documentclass[aps,prd,superscriptaddress,preprint,tightenlines,nofootinbib]{revtex4}

\usepackage{graphicx} 
\usepackage{dcolumn}  
\usepackage{bm}       

\newcommand{\mevcc}{\!\mathrm{MeV}/c^2}

\newcommand{\mev}{\!\mathrm{MeV}}

\newcommand{\gevc}{\!\mathrm{GeV}/c}
\newcommand{\gev}{\!\mathrm{GeV}}

\newcommand{\esigma}{\!\mathrm{\sigma}}

\newcommand{\etac}{\eta_{c}}

\newcommand{\pimp}{\pi^{\mp}}
\newcommand{\pip}{\pi^{+}}
\newcommand{\pim}{\pi^{-}}
\newcommand{\piz}{\pi^{0}}
\newcommand{\kpm}{K^{\pm}}

\newcommand{\kp}{K^{+}}
\newcommand{\km}{K^{-}}
\newcommand{\ks}{K^{0}_{S}}
\newcommand{\jpsi}{J/\psi}
\newcommand{\psip}{\psi(2S)}
\newcommand{\psiboth}{\jpsi~\mathrm{and}~\psip}
\newcommand{\jpsigammaetac}{\jpsi\rightarrow\gamma\etac}
\newcommand{\psipgammaetac}{\psip\rightarrow\gamma\etac}
\newcommand{\psibothgammaetac}{\psiboth\rightarrow\gamma\etac}
\newcommand{\psippipijpsi}{\psip\rightarrow\pip\pim\jpsi}
\newcommand{\brjpsigammaetac}{\mathcal{B}_{1S}}
\newcommand{\brpsipgammaetac}{\mathcal{B}_{2S}}
\newcommand{\brjpsigammaetacfull}{\mathcal{B}(\jpsigammaetac)}
\newcommand{\brpsipgammaetacfull}{\mathcal{B}(\psipgammaetac)}
\newcommand{\brpsippipijpsi}{\mathcal{B}(\psippipijpsi)}

\newcommand{\brratio}{\brjpsigammaetac/\brpsipgammaetac}

\newcommand{\eseven}{E_{\gamma}^{7}}
\newcommand{\ethree}{E_{\gamma}^{3}}
\newcommand{\etwo}{E_{\gamma}^{2}}

\newcommand{\ntwosinc}{\mathrm{N}^{\mathrm{INC}}_{\mathrm{2S}}}
\newcommand{\ntwosexc}{\mathrm{N}^{\mathrm{EXC}}_{\mathrm{2S}}}
\newcommand{\nonesexc}{\mathrm{N}^{\mathrm{EXC}}_{\mathrm{1S}}}
\newcommand{\etwosinc}{\varepsilon^{\mathrm{INC}}_{\mathrm{2S}}}
\newcommand{\etwosexc}{\varepsilon^{\mathrm{EXC}}_{\mathrm{2S}}}
\newcommand{\eonesexc}{\varepsilon^{\mathrm{EXC}}_{\mathrm{1S}}}
\newcommand{\nratio}{\ntwosinc/\ntwosexc}
\newcommand{\eratio}{\eonesexc/\etwosexc}
\newcommand{\bpipi}{\mathcal{B}_{\pi\pi}}
\newcommand{\npsip}{\mathrm{N}_{\psip}}

\newcommand{\ntwosincdata}{59510\pm2145}
\newcommand{\ntwosexcdata}{5376\pm199}
\newcommand{\nratiodata}{11.07\pm0.33}
\newcommand{\nonesexcdata}{5638\pm187}
\newcommand{\etwosincdata}{56.37\%}
\newcommand{\eratiodata}{0.6515}

\newcommand{\brpsipgammaetacdata}{4.32\pm0.16}
\newcommand{\brjpsigammaetacdata}{1.98\pm0.09}
\newcommand{\brratiodata}{4.59\pm0.23}

\newcommand{\brpsipgammaetacsystp}{14}
\newcommand{\brjpsigammaetacsystp}{15}
\newcommand{\brratiosystp}{14}

\newcommand{\brpsipgammaetacsyst}{0.60}
\newcommand{\brjpsigammaetacsyst}{0.30}
\newcommand{\brratiosyst}{0.64}

\newcommand{\psipanswerfull}{(\brpsipgammaetacdata\pm\brpsipgammaetacsyst)\times10^{-3}}

\newcommand{\jpsianswerfull}{(\brjpsigammaetacdata\pm\brjpsigammaetacsyst)\%}

\newcommand{\ratioanswerfull}{\brratiodata\pm\brratiosyst}

\begin{document}

\preprint{CLNS 08/2021}       
\preprint{CLEO 08-05}         

\title{\boldmath $\jpsi$ and $\psip$ Radiative Transitions to $\etac$}

\author{R.~E.~Mitchell}
\author{M.~R.~Shepherd}
\affiliation{Indiana University, Bloomington, Indiana 47405, USA }
\author{D.~Besson}
\affiliation{University of Kansas, Lawrence, Kansas 66045, USA}
\author{T.~K.~Pedlar}
\affiliation{Luther College, Decorah, Iowa 52101, USA}
\author{D.~Cronin-Hennessy}
\author{K.~Y.~Gao}
\author{J.~Hietala}
\author{Y.~Kubota}
\author{T.~Klein}
\author{B.~W.~Lang}
\author{R.~Poling}
\author{A.~W.~Scott}
\author{P.~Zweber}
\affiliation{University of Minnesota, Minneapolis, Minnesota 55455, USA}
% ...
\collaboration{CLEO Collaboration}
\noaffiliation

\date{January 5, 2009}

\begin{abstract}
Using 24.5~million $\psip$ decays collected with the CLEO-c detector at CESR we present the most precise measurements of magnetic dipole transitions in the charmonium system.  We measure
$\brpsipgammaetacfull = \psipanswerfull$,
$\brjpsigammaetacfull / \brpsipgammaetacfull = \ratioanswerfull$, and
$\brjpsigammaetacfull = \jpsianswerfull$.
We observe a distortion in the $\etac$ line shape due to the photon-energy dependence of the magnetic dipole transition rate.  We find that measurements of the $\etac$ mass are sensitive to the line shape, suggesting an explanation for the discrepancy between measurements of the $\etac$ mass in radiative transitions and other production mechanisms.
\end{abstract}

\maketitle


The spectrum of bound charm quarks provides an important testing ground for our understanding of Quantum Chromodynamics (QCD) in the relativistic and non-perturbative regimes.  Radiative transitions in particular have recently been the subject of both lattice QCD calculations~\cite{lqcd} and effective field theory techniques~\cite{eft}.  Key among these are the magnetic dipole (M1) transitions $\jpsigammaetac$ and $\psipgammaetac$, which are among the most poorly measured transitions in the charmonium system.  Not only are precision measurements needed to validate our theoretical understanding, precise measurements of these M1 transitions are critical for normalizing $\etac$ branching fractions, a key input to extracting other properties such as $\Gamma_{\gamma\gamma}(\etac)$.  The $\psibothgammaetac$ transitions are also a source of information on the $\etac$ mass and width.  There is currently a $3.3~\esigma$ inconsistency in previous $\etac$ mass measurements from $\psibothgammaetac$ (averaging $2977.3\pm1.3~\mevcc$) compared to $\gamma\gamma$ or $p\bar{p}$ production (averaging $2982.6\pm1.0~\mevcc$)~\cite{pdg}.


In this Letter, we present the most precise measurements of $\brjpsigammaetacfull$ (abbreviated~$\brjpsigammaetac$), $\brpsipgammaetacfull$ (abbreviated~$\brpsipgammaetac$), and their ratio using 24.5~million $\psip$ decays collected with the CLEO-c detector~\cite{cleodet}.  For the first time, we clearly observe the distortion of the $\etac$ line shape in the photon energy spectrum due to phase space and energy-dependent terms in the M1 transition matrix element.  We find that this line shape distortion may be responsible for the inconsistency in measured $\etac$ mass.


The CLEO-c detector operates at the Cornell Electron Storage Ring (CESR)~\cite{cesr}, which provided symmetric $e^+e^-$ collisions at the $\psip$ center-of-mass.  The detector features a solid angle coverage of $93\%$ for charged and neutral particles. The charged particle tracking system operates in a $1.0~\!\mathrm{T}$ magnetic field along the beam axis and achieves a momentum resolution of $\approx\!0.6\%$ at $p=1~\gevc$. The cesium iodide (CsI) calorimeter attains photon energy resolutions of $2.2\%$ at $E_{\gamma}=1~\gev$ and $5\%$ at $100~\mev$. Two particle identification systems, one based on ionization energy loss ($dE/dx$) in the drift chamber and the other a ring-imaging Cherenkov (RICH) detector, are used together to separate $K^{\pm}$ from $\pi^{\pm}$. Detection efficiencies are determined using a GEANT-based~\cite{geant} Monte Carlo~(MC) detector simulation.


To extract $\brpsipgammaetac$ we use the $\approx~\!\!640~\mev$ photon transition line visible in the inclusive photon energy spectrum from multi-hadronic events collected at the $\psip$ resonance.  A series of exclusive decay modes of the $\etac$ (where the background can be greatly suppressed) are used to constrain the line shape for the inclusive spectrum.  To measure $\brratio$, we take the ratio of events in the chains
\begin{equation}
\psip \rightarrow \pip \pim \jpsi; \; \jpsi \rightarrow \gamma \etac; \; \etac \rightarrow X_i
\label{eq:jpsichain}
\end{equation}
\begin{equation}
\psip \rightarrow \gamma \etac; \; \etac \rightarrow X_i,
\label{eq:psipchain}
\end{equation}
where the $X_i$ are exclusive decay modes of the $\etac$; we then adjust for efficiencies and $\brpsippipijpsi$. Rather than using the inclusive photon spectrum from $\jpsi$ decays, we minimize systematic errors by taking $\brjpsigammaetac$ to be the product of $\brpsipgammaetac$ with $\brratio$.  


The first half of this Letter describes three samples of events: exclusive decays of the $\jpsi$ (Eq.~(\ref{eq:jpsichain})); exclusive decays of the $\psip$ (Eq.~(\ref{eq:psipchain})); and the inclusive photon spectrum from $\psip$ decays.  We use the exclusive samples to investigate the photon-energy dependence of the $\etac$ line shape.  In the second half, our measurement techniques, guided by our line shape investigations, are more fully developed.


Twelve exclusive $\etac$ decay modes are used:  
$X_i$~=~$2(\pip\pim)$,
$3(\pip\pim)$,
$2(\pip\pim\piz)$,
$\pimp\kpm\ks$,
$\piz\kp\km$,
$\pimp\pip\pim\kpm\ks$, 
$\pip\pim\piz\kp\km$,
$2(\kp\km)$,
$2(\pip\pim)\kp\km$,
$\pip\pim\kp\km$,
$\pip\pim\eta$, and
$2(\pip\pim)\eta$,
where the $\eta$ is detected in either its $\gamma\gamma$ or $\pip\pim\piz$ decay mode.
These include all previously reported decay modes of the $\etac$ (except $p\overline{p}$, which has a comparatively small rate)~\cite{pdg} in addition to new decay modes observed here with comparable raw yields.


The reconstruction of the $\psibothgammaetac$ exclusive decay chains share several selection criteria.  In addition to standard fiducial requirements, photons must have energy greater than $30~\mev$ and must not align with the projection of any track into the calorimeter.  For $\piz$ and $\eta$ decays to $\gamma\gamma$, the mass of the pair of daughter photons is required to be within $3~\esigma$ of the nominal mass.  We require fitted tracks of charged particles to have $\chi^2$/d.o.f. $< 50$ and be located in the central region of the detector ($|\cos\theta| < 0.93$, where $\theta$ is the polar angle with respect to the $e^+$ direction).  All charged tracks must be positively identified by a combination of $dE/dx$ and the RICH detector.  To reconstruct $\eta\rightarrow\pip\pim\piz$ all three decay products must pass the above criteria and must have an invariant mass within $30~\mevcc$ of the nominal $\eta$ mass.  The $\ks$ candidates are selected from pairs of oppositely charged and vertex-constrained tracks with invariant mass within $15~\mevcc$ of the $\ks$ mass.  A four-constraint kinematic fit of all identified particles to the initial $\psip$ four-momentum is performed and a $\chi_{4C}^2/$d.o.f $< 5$ is required.  This both sharpens the measured momenta and reduces backgrounds due to missing particles or particle misidentification.  The hypothesis with the best fit quality is accepted per decay mode;  less than 0.5\% of these events enter multiple modes. 


For the selection of exclusive $\jpsi$ decays, the recoil mass of the $\pip\pim$ pair is used to select the process $\psippipijpsi$ and to select $\jpsi$ sidebands.  A sideband subtraction is used to account for non-$\jpsi$ decays.  The $\pip\pim$ recoil momentum is used to boost the photon energy into the $\jpsi$ rest frame.

\begin{figure}
\begin{center}
\includegraphics*[width=\columnwidth]{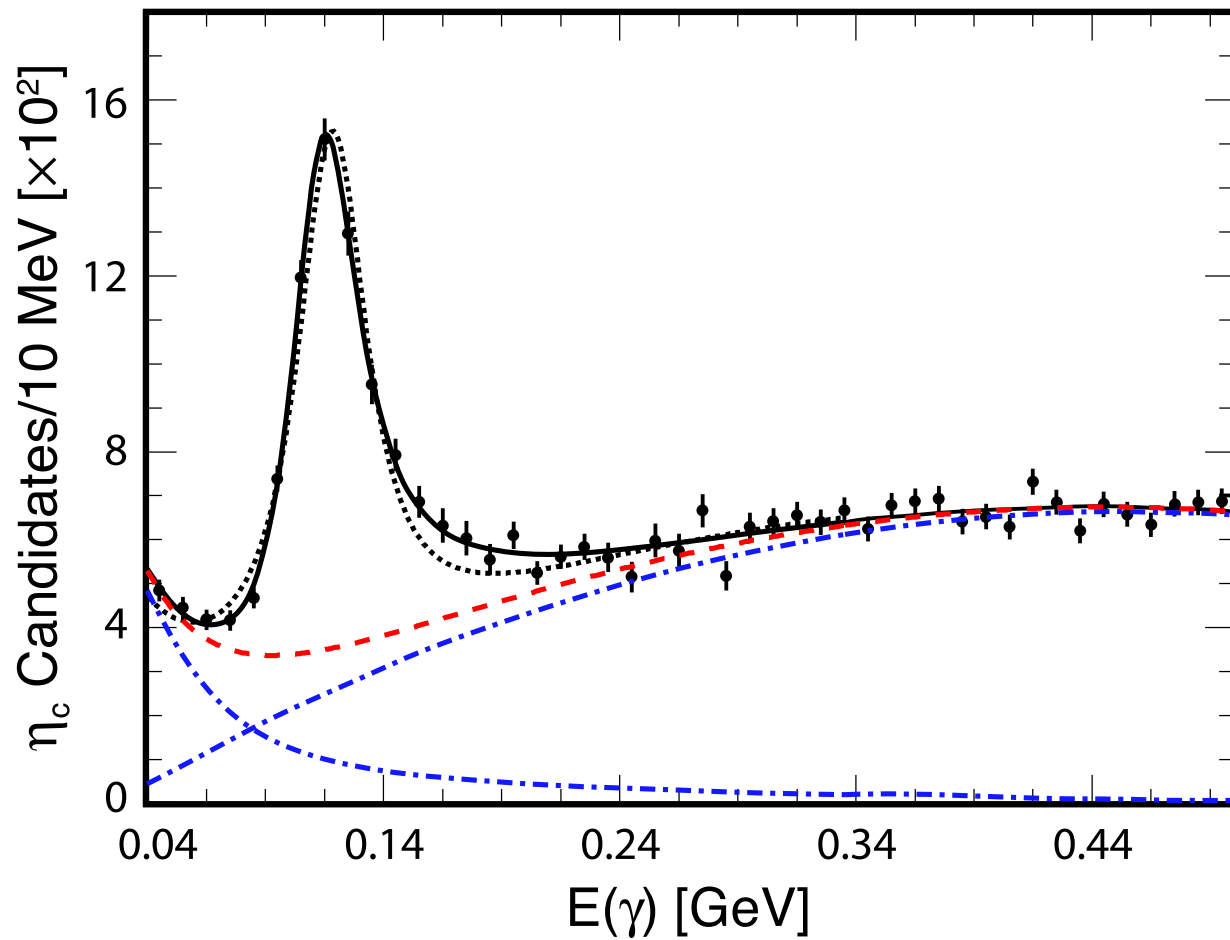}
\caption{ Fits to the photon spectrum in exclusive $\jpsigammaetac$ decays using relativistic Breit-Wigner (dotted) and modified (solid) signal line shapes convolved with a 4.8~$\mev$ wide resolution function.  Total background is given by the dashed line. The dot-dashed curves indicate two major background components described in the text.}
\label{fig:jpsi}
\end{center}
\end{figure}

Fits to the resulting photon energy spectrum for the sum of all $\etac$ decay modes are shown in Fig.~\ref{fig:jpsi}.  The background shape has two essential features:  (i)~background that falls with energy from $\jpsi\rightarrow X_i$, where a spurious cluster is found in the calorimeter, and whose shape is modeled by MC simulation~\cite{istvan} and (ii)~a rising background from both $\jpsi\rightarrow\piz X_i$ and non-signal $\jpsi\rightarrow\gamma X_i$ that is freely fit to a second degree polynomial.  (The polynomial form is motivated by MC and is validated by comparing reconstructed $\jpsi\rightarrow\piz X_i$ in both data and MC.)  These two contributions and their total are shown as dot-dashed and dashed curves.  A fit using an unmodified relativistic Breit-Wigner distribution (dotted line), with the amplitude, mass, and width as free parameters, fails on both the low and high sides of the signal.  A fit using a relativistic Breit-Wigner distribution modified by a factor of $\ethree$~\cite{eft} improves the fit around the peak but leads to a diverging tail at higher energies (not shown).  To damp the $\ethree$ an additional factor of $\mathrm{exp}(-E_{\gamma}^2/\beta^2)$ is added, inspired by the overlap of two ground state wave functions.  The resulting fit has a $25\%$ confidence level (solid line), with $\beta=65.0\pm2.5~\mev$.   In all cases the signal shapes used are convolved with a resolution function determined from MC simulation.  The resolution is $4.8~\mev$ after the kinematic fit.


The uncertainty associated with the line shape prohibits precision mass and width measurements.  It is interesting to note, however, that the resulting $\etac$ mass in the unmodified Breit-Wigner fit is $2976.7\pm0.6~\mevcc$ (statistical error only), consistent with previous measurements from $\psibothgammaetac$, while the modified Breit-Wigner fit returns an $\etac$ mass of $2982.2\pm0.6~\mevcc$ (statistical error only), consistent with that determined from $\gamma\gamma$ fusion and $p\overline{p}$ annihilation.  To resolve this inconsistency, a thorough understanding of the $\etac$ line shape in $\psibothgammaetac$ will be required.

\begin{figure}
\begin{center}
\includegraphics*[width=\columnwidth]{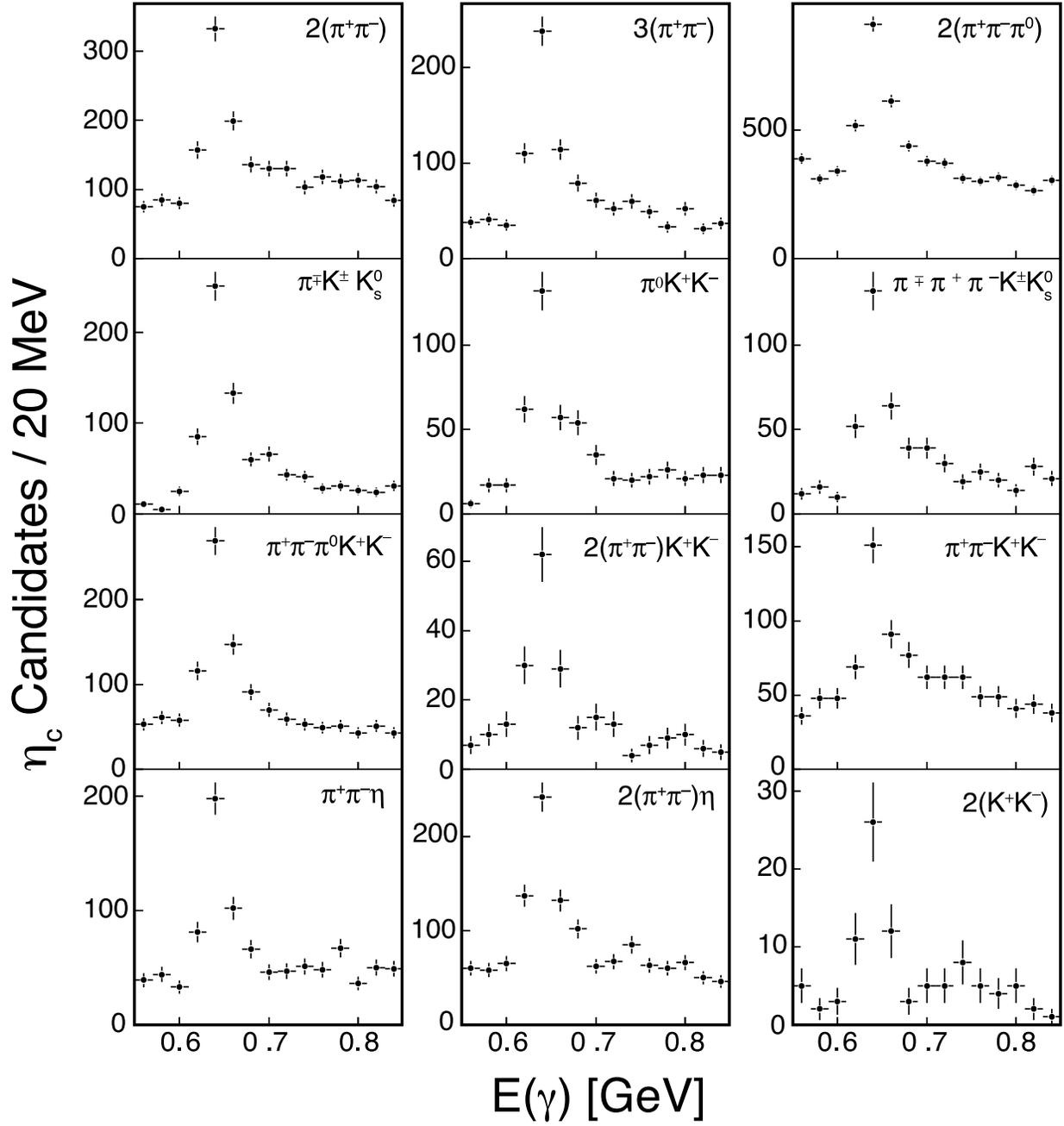}
\caption{The kinematically fitted photon energy spectra from $\psipgammaetac$ for individual $\etac$ decay modes.}
\label{fig:modes}
\end{center}
\end{figure}
\begin{figure}
\begin{center}
\includegraphics*[width=\columnwidth]{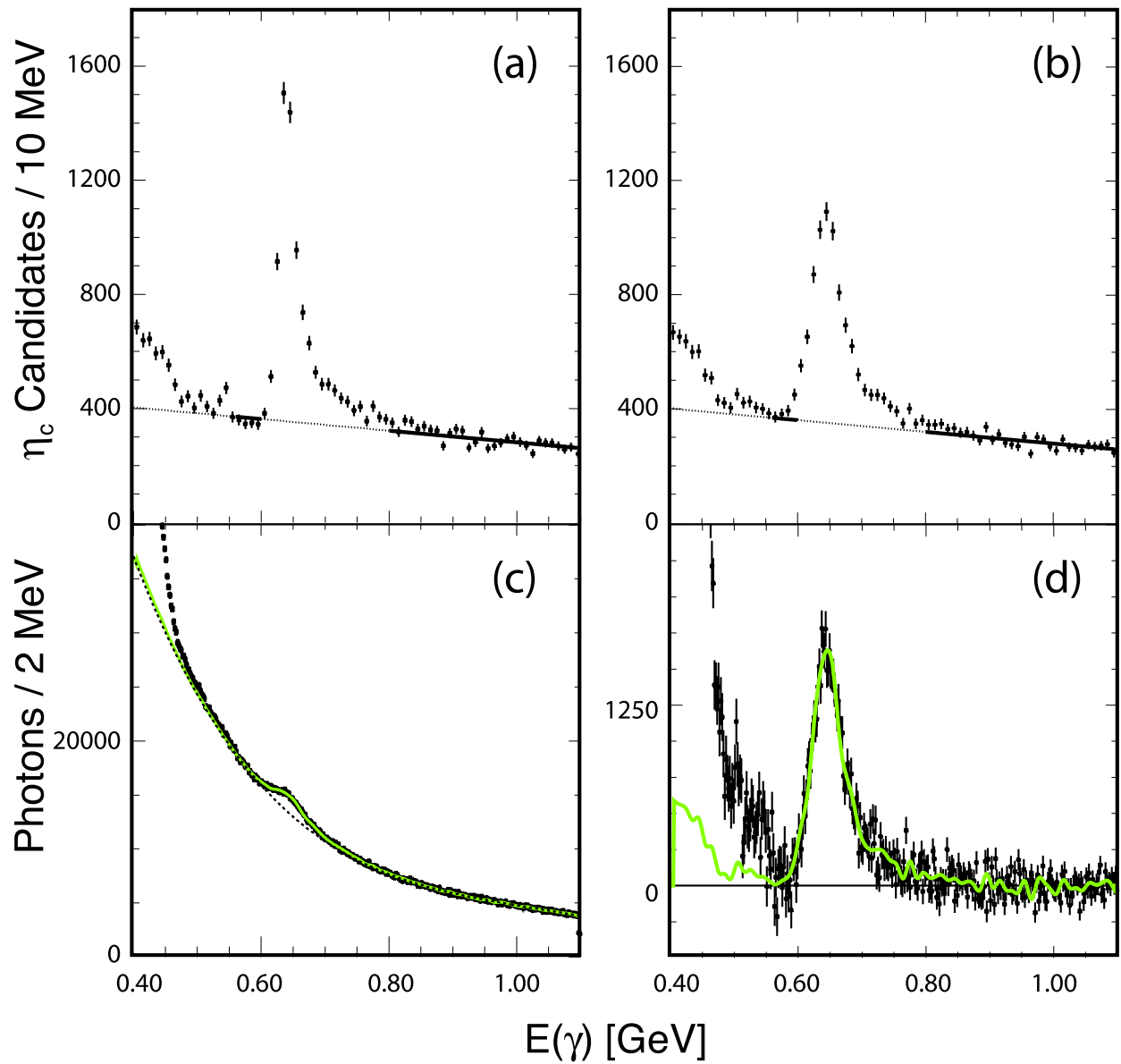}
\caption{ (a)  The kinematically fitted photon spectrum from the sum of exclusive $\psipgammaetac$ modes with a polynomial fit to the background (solid line for the regions included in the fit; dotted line elsewhere).  (b)~The photon energy spectrum that has not been adjusted by the kinematic fit with the same background shape as (a) overlaid.  (c)~The fit to the inclusive photon spectrum in $\psip$ decay.  The signal shape (solid line) is described in the text.  The background is given by the dashed line.  (d)~The background subtracted inclusive photon spectrum with the signal shape overlaid (solid line).}
\label{fig:psip}
\end{center}
\end{figure}

For the selection of exclusive $\psip$ decays, the transition $\psippipijpsi$ is suppressed by requiring there be no pair of oppositely charged particles (assumed to be pions) with recoil mass within $15~\mevcc$ of the $\jpsi$ mass.  The photon energy spectra for individual $\etac$ decay modes are shown in Fig.~\ref{fig:modes}; the sum of all modes is shown in Fig.~\ref{fig:psip}a.  Several small nonlinear backgrounds below $560~\mev$ are apparent and are due to a combination of (i)~$\psip\rightarrow\piz h_c; h_c\rightarrow\gamma\etac$; (ii)~$\psip\rightarrow\gamma\chi_{cJ}; \chi_{cJ}\rightarrow\gamma\jpsi$; and (iii)~$\psip\rightarrow\piz\jpsi$.  Based on detailed MC studies, all other backgrounds are linear, the largest being $\psip\rightarrow\piz X_i$.


Fits to the $\psipgammaetac$ photon energy spectrum with a relativistic Breit-Wigner distribution convolved with an experimental resolution function (with a resolution of $5.1~\mev$ after the kinematic fit) were unsuccessful.  For a hindered M1~transition the matrix element acquires terms proportional to $\etwo$, which, when combined with the usual $\ethree$ term for the allowed transitions, lead to contributions in the radiative width proportional to $\eseven$~\cite{eft}. We find that if we assume a linear background, as indicated by MC simulations, we are not able to obtain a good fit to our $E_{\gamma}$ spectrum for the sum of exclusive $\psipgammaetac$ modes with a pure $\eseven$ dependence. We therefore use the empirical procedure described below to extract the $\psipgammaetac$ yield.


Extensive cross-checks have been performed to verify that the line shape asymmetry is not an experimental artifact.  Events selected without the aid of a kinematic fit indicate an asymmetric line shape independently in both the photon energy and the hadronic mass.  The asymmetric line shape is not correlated with $\etac$ decay modes that include $\piz$,  $\ks$, or $\eta$ candidates.  No indication of either asymmetry or peaking background has been found in detailed MC studies, where all known decays in the charmonium and light quark systems are simulated and unknown decays are modeled with the EvtGen generator~\cite{evtgen}. The photon angular distribution from $\psipgammaetac$ fits the $1+\cos^2\theta$ distribution expected for M1~transitions, whether using symmetric or asymmetric signal shapes to extract yields in different regions of $\cos\theta$.
 

For the selection of the third sample of events, the inclusive photon spectrum from $\psip$ decays, the photon is required to pass the same requirements and $\psippipijpsi$ is suppressed in the same manner as the exclusive $\psipgammaetac$.  To suppress the $\piz$ background, each signal photon candidate is paired with all other photons in the event and is rejected if the invariant mass of the pair is within three standard deviations of the $\piz$ mass.  Backgrounds due to $e^+e^-$ QED processes are substantially reduced by requiring one or more charged tracks in an event in combination with requirements on total energy and track momenta~\cite{cleom1}.  The inclusive photon spectrum is shown in Fig.~\ref{fig:psip}c. The rise at lower energies is from $\chi_{cJ}\rightarrow\gamma\jpsi$.  All other backgrounds are smooth and are dominated by $\piz$ decays.


Our final results are obtained from:
\begin{equation}
\brpsipgammaetac = \frac{\ntwosinc}{\etwosinc\npsip}
\end{equation}
\begin{equation}
\frac{\brjpsigammaetac}{\brpsipgammaetac} = \frac{\nonesexc}{\ntwosexc\left( \eratio \right){\bpipi}}
\end{equation}
\begin{equation}
\brjpsigammaetac = \frac{\left( \nratio \right) \nonesexc}{\etwosinc \left( \eratio \right){\npsip\bpipi}}
\end{equation}
where $N$ and $\varepsilon$ represent the observed yields and the calculated efficiencies of $\psip$ and $\jpsi$ in inclusive (INC) and exclusive (EXC) $\etac$ channels.  The $\brpsippipijpsi$, abbreviated $\bpipi$, is taken from a previous CLEO measurement, $(35.04\pm0.07\pm0.77)\%$~\cite{bpipi}.  $\npsip$ is the number of $\psip$ decays, 24.5~million, which is known to $2\%$~\cite{bpipi}.    Final values are listed in Table~\ref{tab:results}.  Systematic errors are listed in Table~\ref{tab:systematics}.

\begin{table}
\caption{Final yields and efficiencies.}
\label{tab:results}
\medskip
\begin{center}
\begin{tabular}{ll}
\hline
\hline
$\ntwosinc$ & $\ntwosincdata$ \\
$\ntwosexc$ &$\ntwosexcdata$ \\
$\nratio$ &$\nratiodata$ \\
$\nonesexc$& $\nonesexcdata$ \\
$\etwosinc$ &$\etwosincdata$ \\
$\eratio$ &$\eratiodata$ \\
\hline
\hline
\end{tabular}
\end{center}
\end{table}

\begin{table}
\caption{Summary of systematic errors (in percent).}
\label{tab:systematics}
\medskip
\begin{center}
\begin{tabular}{lccc}
\hline
\hline
Systematic Error (\%)  &  $\brpsipgammaetac$  &  $\brjpsigammaetac$ & $\brratio$ \\
\hline
Fitting for $\ntwosinc$             & 8 & -- & -- \\
Fitting for $\ntwosexc$             & -- & -- & 8 \\
Fitting for $\nratio$             & -- & 4 & -- \\
Fitting for $\nonesexc$             & -- & 10 & 10 \\
Effect of Line Shape on $\etwosinc$ and $\etwosexc$			& 7 & 4 & 3 \\
Effect of Line Shape on $\eonesexc$ 			& -- & 1 & 1 \\
MC Modeling of Inclusive $\etac$  Decays  &&& \\ 
~~~and Inclusive Event Selection       	   & 8 & 8 & -- \\
Exclusive Efficiency Ratio Calculation      & -- & 2 & 2 \\
Exclusive Event Selection    	 & -- & 3 & 3 \\
$640~\mev$ Photon Efficiency     & 2 & -- & 2 \\
$110~\mev$ Photon Efficiency     & -- & 2 & 2 \\
$\pip\pim$ Efficiency    	 & -- & 2 & 2 \\
$\npsip$    			 & 2 & 2 & -- \\
$\brpsippipijpsi$     		& -- & 2 & 2 \\
{\bf Total }                     & {\bf \brpsipgammaetacsystp } & {\bf \brjpsigammaetacsystp } & {\bf \brratiosystp } \\
\hline
\hline
\end{tabular}
\end{center}
\end{table}


The procedure used to obtain $\ntwosinc$, $\ntwosexc$, and $\nratio$ is threefold:  (i)~the background to the exclusive $\psipgammaetac$ process is fit with a first order polynomial using regions above and below the signal (Fig.~\ref{fig:psip}a);  (ii)~the background is then fixed and carried to the exclusive $\psipgammaetac$ spectrum that has not been adjusted by a kinematic fit (Fig.~\ref{fig:psip}b), which is directly comparable to the inclusive spectrum;  (iii)~the histogram obtained by subtracting off the background in (ii) is used to fit (along with a fourth degree polynomial for the background) the inclusive photon spectrum (Fig.~\ref{fig:psip}c).  Numbers of events above background are counted for photon energies between $560~\mev$ and $1100~\mev$.  As can be seen from the background subtracted inclusive spectrum (Fig.~\ref{fig:psip}d), there is excellent agreement between the exclusively determined signal shape (line) and the signal shape present in the inclusive photon spectrum.


Systematic errors due to this fit procedure are determined by varying the orders of the background polynomials, the ranges of the fits, and the range of the signal that is excluded in the exclusive fit.  We find $8\%$ variations for $\ntwosinc$ and $\ntwosexc$, but only $4\%$ variations in their ratio.  In addition, since we integrate our signal between $560~\mev$ and $1100~\mev$, our uncertainty in the line shape affects the signal efficiency.  We take a conservative systematic error of $7\%$ for $\etwosinc$ to cover the two extreme cases that the signal shape has a Breit-Wigner tail and the case that the entire signal lands within the signal region.  The systematic error is smaller for $\etwosexc$ ($3\%$) because the kinematic fit pulls more of the signal within the signal region.  These errors are correlated resulting in an error of $4\%$ in the ratio.


The measurement of $\nonesexc$ is from the modified Breit-Wigner fit shown in Fig.~\ref{fig:jpsi}, which results in an $\etac$ width of $31.5\pm1.5~\mevcc$ (statistical error only).   A fit using an unmodified Breit-Wigner distribution with the $\etac$ width fixed to the current PDG value ($26.5~\mevcc$) gives a smaller value of $\nonesexc$ by $10\%$.  A $10\%$ systematic error covers this extreme minimum, as well as variations using a Breit-Wigner distribution modified only by an $\ethree$ term and variations of the background parametrization.  The signal is integrated above $40~\mev$, which introduces an additional $1\%$ systematic error into the efficiency estimate ($\eonesexc$) due to signal shape uncertainty.  All efficiences are independent of photon energy in the signal regions; hence the signal shape uncertainty only affects efficiencies by way of the limits of integration.


Because only $\approx\!25\%$ of all $\etac$ decays are known, our estimate of $\etwosinc$ depends on modeling the unknown decays of the $\etac$.  We use two different hadronization models, but then weight the efficiencies for different track multiplicities according to multiplicities observed in data.  We account for our limited sensitivity to events with no tracks by varying their weight by $\pm100\%$.  We also simultaneously vary the event selection requirements, keeping and removing the $\piz$ suppression, and using several QED suppression schemes.  All variations are covered with an $8\%$ systematic error.


The ratio of exclusive efficiencies, $\eratio$, is calculated by weighting the efficiency ratio for each $\etac$ decay mode individually using the number of $\psipgammaetac$ events observed in each mode.  The ratio of efficiencies has a slight dependence on decay mode.  Varying the number of events in each mode by $30\%$, we find a $2\%$ systematic error due to the composition of decay modes.


Many systematic errors, such as those due to tracking or $\piz$ reconstruction efficiencies, cancel in the ratio of exclusive efficiencies, $\eratio$.  To estimate any possible dependence of the final numbers on the exclusive event selection, a wide range of requirements on the quality of the kinematic fit are imposed, and photons are required to be in different regions of the calorimeter, resulting in variations of less than $3\%$.  We conservatively assume that systematic uncertainties for the reconstruction efficiencies of the $\approx\!110~\mev$ and $\approx\!640~\mev$ transition photons do not cancel in the ratio and assign $2\%$ uncertainties for each.  We also assign a $2\%$ uncertainty for the efficiency of the $\pip\pim$ from $\psippipijpsi$.


We find 
$\brpsipgammaetac = \psipanswerfull$,
$\brjpsigammaetac / \brpsipgammaetac = \ratioanswerfull$, and
$\brjpsigammaetac = \jpsianswerfull$.
Both M1 transitions $\brjpsigammaetac$ and $\brpsipgammaetac$ are larger than previous measurements~\cite{cleom1,cbm1} (56\% and 44\% higher than the PDG averages~\cite{pdg}, respectively) due to a combination of a larger $\etac$ width and an accounting for the asymmetry in the line shapes.

In conclusion, we have studied $\psibothgammaetac$ transitions and measured their branching fractions.  These new measurements will renormalize many $\etac$ branching fractions and $\Gamma_{\gamma\gamma}(\etac)$.  We also find that a thorough theoretical understanding of the $\etac$ line shape in M1 transitions in the charmonium system will be crucial if the systematic errors on these branching fractions are to be reduced and if the mass and width of the $\etac$ are to be extracted from these processes.

We gratefully acknowledge the effort of the CESR staff
in providing us with excellent luminosity and running conditions.  We also thank Jozef Dudek and Nora Brambilla for useful discussions.  
This work was supported by 
the A.P.~Sloan Foundation, 
the National Science Foundation, 
the U.S. Department of Energy, 
the Natural Sciences and Engineering Research Council of Canada, and 
the U.K. Science and Technology Facilities Council.

\newpage

\newpage

\newpage

\end{document}